\documentclass[aps,pra,twocolumn,showpacs,superscriptaddress,longbibliography,amsmath,amssymb,floatfix]{revtex4-1}

\usepackage[dvips]{graphicx}

\usepackage{dcolumn}
\usepackage{bm}
\usepackage{ulem}
\newcommand{\vect}[1]{\mathbf{#1}}

\begin{document}

\title{Electron spin resonance in spiral antiferromagnet linarite: theory and experiment}

\author{S. K. Gotovko}
\affiliation{P. L. Kapitza Institute for Physical Problems, Russian Academy of Sciences, 119334 Moscow, Russia}
\affiliation{National Research University Higher School of Economics, 101000 Moscow, Russia}

\author{L. E. Svistov}
\email{svistov@kapitza.ras.ru}
\affiliation{P. L. Kapitza Institute for Physical Problems, Russian Academy of Sciences,  119334 Moscow, Russia}

\author{A. M. Kuzmenko}
\affiliation{Prokhorov General Physics Institute, Russian Academy of Sciences, 119991 Moscow, Russia}

\author{A. Pimenov}
\affiliation{Institute of Solid State Physics, Vienna University of Technology, 1040 Vienna, Austria}

\author{M. E. Zhitomirsky}
\email{mike.zhitomirsky@cea.fr}
\affiliation{Universit\'e Grenoble Alpes, CEA, INAC-PHELIQS F-38000 Grenoble, France}

\date{\today}

\begin{abstract}
We present combined experimental and theoretical investigation of the low-frequency ESR 
dynamics in the ordered phases of magnetic mineral linarite. This material consists of weakly 
coupled spin-1/2 chains of copper ions with frustrated ferro- and antiferromagnetic interactions. 
In zero magnetic field, linarite orders into a spiral structure and exhibits a peculiar magnetic phase 
diagram sensitive to the field orientation. The resonance frequencies and their field dependence 
are analyzed combining microscopic and macroscopic theoretical approaches and precise values 
of magnetic anisotropy constants are obtained. We conclude that possible realization of exotic 
multipolar quantum states in this material is greatly influenced by the biaxial anisotropy.
\end{abstract}

\pacs{ }

\maketitle

\section{Introduction}

The frustrated spin-1/2 chain model with nearest neighbor ferromagnetic $J_1<0$ and next-nearest 
neighbor antiferromagnetic $J_2>0$ exchanges has recently attracted a great deal of  interest 
owing to its exotic quantum properties. In strong magnetic fields the model can exhibit the longitudinal 
spin-density wave, the spin nematic, and even higher-order multipolar phases
\cite{Chubukov91,Heidrich06,Kecke07,Hikihara08,Sudan09,Heidrich09,Shindou09,Zhitomirsky10,Nishimoto15}.
Frustrated ferromagnetic chains are realized in a family of copper-oxide materials with edge-sharing 
CuO$_2$ plaquettes represented, for instance, by LiCuVO$_4$ \cite{Gibson04,Enderle05}, 
Rb$_2$Cu$_2$Mo$_3$O$_{12}$ \cite{Hase_2004}, LiCu$_2$O$_2$ \cite{Matsuda_2004}, 
NaCu$_2$O$_2$ \cite{Drechsler_2006}, Li$_2$ZrCuO$_4$ \cite{Drechsler_2007} and 
PbCuSO$_4$(OH)$_2$ (linarite) \cite{Baran_2006}. 
Among these the natural mineral linarite PbCuSO$_4$(OH)$_2$ 
\cite{Baran_2006,Yasui_2011,Willenberg_2012,Wolter_2012,Schapers_2013,Willenberg_2016,
Povarov_2016,Rule_2017,Cemal_2018,Feng18,Heinze19}
combines a moderate saturation field of about 10~T with close proximity to the quantum critical point 
$|J_2/J_1|_c = 1/4$, which may provide direct access to the most exotic multipolar states \cite{Hikihara08}.
Besides that, linarite exhibits a unique phase diagram for magnetic fields applied along the chain direction
with up to five  commensurate and incommensurate phases
\cite{Willenberg_2012,Willenberg_2016,Feng18}. 

There is an ongoing debate on the role of anisotropy for the observed properties of linarite
\cite{Cemal_2018,Feng18,Heinze19}. Indeed, the phase diagram changes dramatically once magnetic 
field is tilted away from the chain direction. Cemal {\it et al.}~\cite{Cemal_2018} have suggested
the minimal anisotropic model with orthorhombic symmetry and estimated corresponding
microscopic parameters for linarite on the basis of the inelastic neutron scattering measurements in the 
high-field polarized phase. Here, we present our experimental results on the low-frequency dynamics in linarite 
studied by the electron spin resonance (ESR) technique together with theoretical analysis
based on the microscopic model as well as on the phenomenological  hydrodynamic theory.
The main advantage of the ESR method for studying the anisotropy effects is its high frequency/energy resolution,
which allows us to obtain much more reliable values of the anisotropy constants.
In the remaining part of Introduction we review the basic crystallographic and magnetic properties 
of linarite that are important for our subsequent analysis.

\begin{figure}[t]
\includegraphics[width=0.9\columnwidth,angle=0,clip]{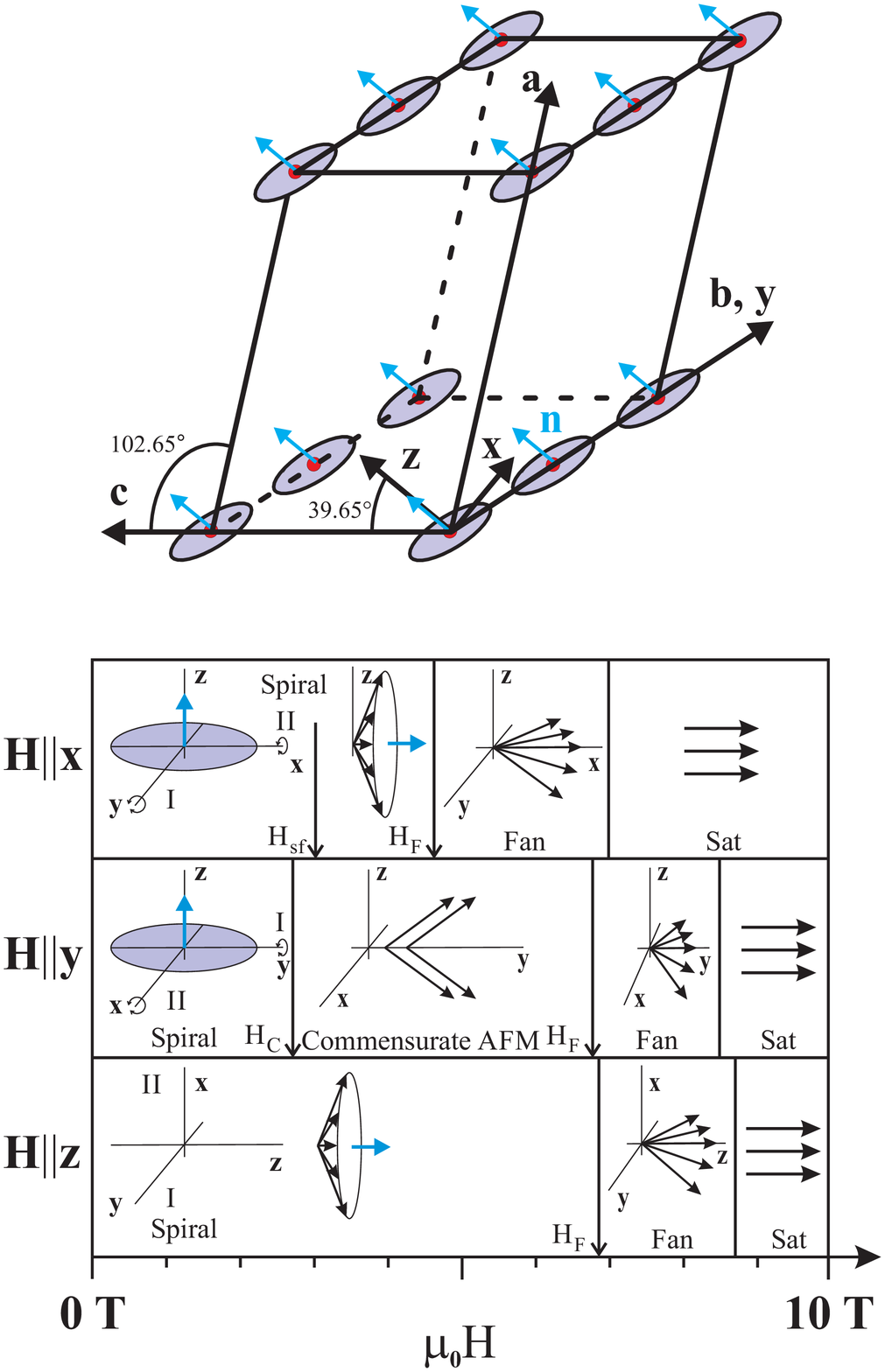}
\caption{(Color online.) Upper panel: 
Schematic crystal structure of linarite showing the positions of magnetic copper ions (small red circles). 
The spiral magnetic structure in zero field is illustrated by grey ellipses. 
The orthogonal triad $\hat{\bf x},\hat{\bf y},\hat{\bf z}$ marks the principal axes of the anisotropy tensor.
The hard axis $\hat{\bf z}$ is perpendicular to the spiral plane, whereas the intermediate axis $\hat{\bf y} \parallel{\bf b}$.
Bottom panel: Magnetic structures appearing for different orientations
of an external field for $T=1.3$~K.}
\label{structurenew}
\end{figure}

The crystal lattice of linarite belongs to the monoclinic space group $P2_1/m$ with lattice parameters 
$a = 9.70$~\AA, $b = 5.65$~\AA, $c = 4.69$~\AA,  and $\beta = 102.7^{\circ}$ 
\cite{Effenberger_1987}. Figure~\ref{structurenew} shows a schematic crystal structure including
only  Cu$^{2+}$ ions. The CuO$_2$ plaquettes form weakly pleated ribbons and  
the crystal unit cell contains two adjacent copper ions along the $b$ axis. 
Still, from the point of view of isotropic exchange interactions, the copper chains 
remain uniform with the same exchange coupling $J_1$ in all spin pairs at distance $b/2$
and $J_2$ for second nearest neighbors at distance $b$.
In zero field, linarite magnetically orders at $T_N \approx 2.8$~K into an elliptic spiral structure
with the propagation vector ${\bf k}_{ic} = (0, 0.189, 1/2)$ in the reciprocal lattice units \cite{Willenberg_2012}.
The spin spiral rotates in the $xy$ plane, where the $y$ axis is parallel to the crystallographic $b$ axis and the $x$ axis lies in the $ac$ plane making  an angle 27$^\circ$ with the $a$ axis. Two components of the order parameter
in this elliptic spiral state are  
$\mu_x\approx 0.64~\mu_B$ and $\mu_y\approx 0.83~\mu_B$ \cite{Willenberg_2012}.

In applied field, linarite exhibits a variety of magnetic structures depending on field strength and orientation, 
see Fig.~\ref{structurenew}. Magnetic phases and transition fields presented in Fig.~\ref{structurenew} are
given in accordance with Refs.~\cite{Cemal_2018, Povarov_2016}. For magnetic field applied parallel to the spiral 
plane  along the easy $x$ axis, the observed phase sequence, spin helix --- spin cone --- spin fan, conforms to the one expected for magnetic spirals in the presence of anisotropy \cite{Nagamiya67}.  For the orthogonal in-plane direction 
${\bf H}\parallel{\bf b}$, an additional commensurate phase characterized by ${\bf k}_{c} = (0, 0, 1/2)$ 
appears instead of the conical state in a wide range of fields. Its presence has been attributed to the biaxial anisotropy, which competes  with an incommensurate tendency set by competing exchanges \cite{Cemal_2018}. 
Common to all field directions is the fan phase stabilized  in the vicinity of the transition into the 
polarized state as expected on general symmetry arguments \cite{Nagamiya67}. 
An alternative scenario  put forward in \cite{Willenberg_2016}
identifies the  high-field phase with the longitudinal 
spin-density wave (SDW) characteristic to  one-dimensional $J_1$--$J_2$ chains. 
Such a suggestion is motivated by observation of 
a peculiar field/temperature dependence of the ordering wave vector in high fields. However,
Cemal {\it et al.} \cite{Cemal_2018} have subsequently shown that
the nontrivial field dependence $k_{ic}(H)$ at the lowest temperature $T= 60$~mK is at least
partially accounted for by the effect of magnetic anisotropy in the fan phase.
In any case, at the transition to the saturated phase the ordering wave vector $k_{ic}$ remains finite in linarite \cite{Willenberg_2016,Cemal_2018}
in contradiction to the SDW scenario, which predicts   $k_{ic}\to 0$.

The recent high-field NMR measurements for ${\bf H}\parallel{\bf b}$
leave a narrow window of fields 9.35~T~$\leq H\leq 9.64$~T, where a spin nematic phase
may be present  in linarite \cite{Heinze19}. A rather small field region for the exotic quantum phase 
is not entirely surprising in view of relatively large ordered moments in this material in zero magnetic field 
and, consequently, small quantum fluctuations.
In contrast,  LiCuVO$_4$, another candidate material for the high-field nematic phase
\cite{Svistov11,Orlova17}, has notably smaller moments $\sim 0.3\mu_B$ \cite{Gibson04}
making it a more feasible  venue for observing exotic quantum physics albeit in higher
magnetic fields.
Note, that even in the absence of the multipolar quantum states the phase diagram of linarite 
provides an interesting example of an incommensurate magnet under combined effect of anisotropy and
magnetic field  \cite{Nagamiya67}.

The paper is organized as follows. In Section II we provide the theoretical  basis for interpreting the low-frequency ESR dynamics using both the microscopic spin-wave approach (Sec.~IIA) 
and the macroscopic field-theoretical treatment (Sec.~IIB). Section~III describes experimental 
results that are compared to the theoretical predictions. 
In Sec.~IV we conclude by emphasizing the main consequences to the physics of linarite.

\section{Theory}
\label{sec:Theory}

\subsection{Spin Model}
\label{sec:model}

In accordance with the monoclinic symmetry of linarite crystals, we base our theoretical consideration 
on a general anisotropic exchange Hamiltonian for spins $S=1/2$ written in the global coordinate frame as 
\begin{equation}
\hat{\cal H} = \sum_{\langle ij\rangle} \bigl[
J_{ij}^{xx} S_i^x S_j^x + J_{ij}^{yy} S_i^y S_j^y + J_{ij}^{zz} S_i^z S_j^z \bigr],
\label{H0}
\end{equation}
where $\hat{\bf y}$  is chosen along the two-fold $b$ axis and $\hat{\bf x}$ and $\hat{\bf z}$ 
are oriented in the $ac$ plane. The Hamiltonian (\ref{H0}) provides the minimal anisotropic spin model 
for linarite. In particular, we assume that principal axes of the symmetric exchange tensor are the same 
for each bond, which is generally not true. However, as we shall see below, the exchange anisotropies 
contribute additively to the ESR gaps. Therefore, in the case of a large difference between exchange 
bonds, one can take into account the anisotropy of the strongest bond only. Also, the pleated structure 
of the CuO$_2$ ribbons in linarite allows for the antisymmetric Dzyaloshinskii-Moriya (DM) interaction 
on the nearest-neighbor bonds. We omit this interaction in the minimal model since 
(i) due to staggering of the DM vectors it does not affect the pitch angle of the spin spiral and 
(ii) the inelastic neutron scattering (INS) measurements \cite{Rule_2017,Cemal_2018} 
successfully fit  the magnon dispersion data without the  DM term.

There is an emergent consensus in literature that magnetic properties of linarite are determined by
competing ferromagnetic nearest-neighbor $J_1$ and antiferromagnetic second-neighbor $J_2$ exchanges 
inside the spin chains as well as by an interchain coupling $J_{ic}$. Still, different authors proposed
quite different values of the microscopic parameters, see Table~\ref{parameters},
where we also indicated the experimental techniques used in each study.
An accurate estimate of the exchange parameters of linarite was obtained by the INS measurements 
in high magnetic fields \cite{Cemal_2018}, which suppress quantum fluctuations.
In the same work, the dominant interchain coupling was identified as  $J_c$, exchange between nearest 
neighbors in the $c$ direction. In contrast, the zero-field neutron measurements  \cite{Rule_2017} suggested a dominant diagonal interchain coupling $J_c'$ (quoted in Table~\ref{parameters}).  
This conclusion was based on the spin-wave fits of the low-energy part
of the spectrum, where according to the dynamical DMRG simulations  \cite{Sirker12}
quantum renormalization effects are significant and the harmonic spin-wave theory is poorly applicable.

\begin{table}
\caption{\label{parameters} 
Microscopic exchange constants of linarite
from different works together with
the computed ratio of magnetic susceptibilities 
using Eqs.~(\ref{chi_per}) and (\ref{chi_par}).
The last two rows give 
$\chi_{\perp}/\chi_{\parallel}$ measured experimentally   
\cite{Yasui_2011} or directly derived from our ESR data.}
\begin{center}
\begin{tabular}{ccccc}
\hline
\hline
 Method & $J_1$ [K] & $J_2$ [K] & $J_{ic}$ [K] & $\chi_{\perp}/\chi_{\parallel}$ \\
\hline
& & & & \\
$\chi(T)$ \cite{Yasui_2011} & $-13\pm 3$ & $21\pm 5$ & $-$ & $2.7$ \\
$M(H)$, $\chi(T)$ \cite{Wolter_2012} & $-100$ & 36 & $-$ & $6.7$ \\
INS ($H=0$) \cite{Rule_2017}  & &  &  &  \\
LSWT   & $-114 \pm 2$ & $37 \pm 1$ & $4 \pm 0.5$ & 2.2 \\
DMRG  & $-78$ & $28$ & $7$ & 1.7 \\
INS ($H>H_{\rm sat}$) \cite{Cemal_2018} & $-168$ & $46$ & $8$ & 2.2 \\[1mm]
\hline \\[-2mm]
$M(H)$, $T = 2$~K \cite{Yasui_2011} &  &  & & $1.6 \pm 0.1$ \\
ESR, $T=1.3$~K  &  &  &  & $1.85 \pm 0.1$ \\
(this study) &  &  &  &  \\
\hline
\hline
\end{tabular}
\end{center}
\end{table}

In the rest of the article, both $J_2$ and $J_c$ are assumed to be isotropic, whereas the biaxial anisotropy 
of the ferromagnetic nearest-neighbor bonds is represented as
\begin{equation}
\hat{\cal H} = J_1 \sum_{\langle ij\rangle} \bigl[ (1+\varepsilon) S_i^x S_j^x + (1-\varepsilon) 
S_i^y S_j^y + (1-\delta)  S_i^z S_j^z \bigr]
\label{Ha}
\end{equation}
with $0 < \varepsilon < \delta$ such that  $x$ and $z$ are the easy and the hard axis, respectively.
The same choice of axes was adopted by Cemal {\it et al.} \cite{Cemal_2018}, though our definitions
of $\delta$ and $\varepsilon$ are somewhat different.

\subsection{Microscopic Theory}
\label{sec:microTh}

The spin-wave theory  of incommensurate helical magnetic structures was developed 
by various authors in the sixties \cite{Cooper62,Cooper63,Elliot66,Nagamiya67,Cooper68}, 
see also recent works \cite{Zhitomirsky96,Chen13,Milstein15}. 
The motivation for the early studies  was chiefly from the experimental investigation 
of rare-earth compounds dominated by the single-ion anisotropy. 
In this section, we extend the previous
analysis to the case of the symmetric exchange anisotropy relevant for Kramers ions with
an effective spin $S=1/2$ and provide a few additional results.

\subsubsection{Zero magnetic field}

We assume that competing exchange interactions  produce a spiral 
spin structure in the $xy$ plane. The first standard step consists in 
transformation from the fixed global frame to the rotating local coordinate axes $(x_i,y_i,z_i)$ such that 
$\hat{\bf z}_i$ is always oriented along the equilibrium spin direction on a given site $i$ and  
$\hat{\bf y}_i$ is orthogonal
to the plane of the spiral. Spin components in the global frame 
(denoted below by $0$ subscripts) are related to those in the rotating local frame  by
\begin{eqnarray}
&& S_i^{x_0} = S_i^z \cos\theta_i  - S_i^x \sin\theta_i \,, 
\label{LF}\\ 
&& S_i^{y_0} = S_i^z \sin\theta_i  + S_i^x \cos\theta_i \,, \quad S_i^{z_0}=S_i^y\,,
\nonumber 
\end{eqnarray}
where $\theta_i$ is the rotation angle to be determined later.
It is convenient to introduce
\begin{equation}
J_{ij} = \frac{1}{2} \bigl(J_{ij}^{xx} + J_{ij}^{yy}\bigr)\,, \quad 
\varepsilon_{ij} = \frac{1}{2} \bigl(J_{ij}^{xx} - J_{ij}^{yy}\bigr)\,.
\end{equation}
The Hamiltonian (\ref{H0}) written in the local frame becomes
\begin{eqnarray}
\hat{\cal H} & = & \sum_{\langle ij\rangle} \Bigl[
J_{ij} \cos(\theta_i-\theta_j) (S_i^z S_j^z + S_i^x S_j^x)  + J_{ij}^{zz} S_i^y S_j^y 
\nonumber \\
& & \mbox{}+ \varepsilon_{ij} \cos(\theta_i+\theta_j) (S_i^z S_j^z - S_i^x S_j^x) + \ldots
 \Bigr],
\label{Hloc0}
\end{eqnarray}
In the above expression we dropped mixed terms like $S_i^zS_j^x$ since those play no role in the following
calculations. 

Zero-temperature classical energy is obtained from Eq.~(\ref{Hloc0}) by neglecting all fluctuations, 
$S_i^z\to S$, $S_i^{x,y}\to 0$,
\begin{equation}
E_{\rm cl}  = S^2\sum_{\langle ij\rangle}  \bigr[
J_{ij} \cos(\theta_i-\theta_j) +\varepsilon_{ij} \cos(\theta_i+\theta_j) 
 \bigr].
\label{Ecl0}
\end{equation}
For uniaxial planar anisotropy ($\varepsilon_{ij}=0$), spins rotate uniformly in space by $\theta_i = {\bf Q}\cdot{\bf r}_i$,
where $\bf Q$ corresponds to the minimum of the Fourier transform
\begin{equation}
J_{\bf q} = \sum_j J_{ij}\, e^{i{\bf q}({\bf r}_i-{\bf r}_j)} \ .
\label{Jq}
\end{equation}
For the microscopic model of linarite we have
\begin{equation}
J_{\bf q} =  -2|J_1|\cos q_y + 2J_2 \cos 2q_y + 2J_c \cos q_z\ ,
\end{equation}
where we set all nearest-neighbor distances to 1.
The minimum is achieved for ${\bf Q} = (0,Q,\pi)$ with
\begin{equation}
\cos Q  = \frac{|J_1|}{4J_2} \  .
\label{cosQ}
\end{equation}
A sizeable second-neighbor exchange  $J_2 \sim 0.3 |J_1|$  produces the incommensurate 
spin spiral along the copper chains, whereas $J_c>0$ is responsible
for antiferromagnetic spin arrangement between chains in the $c$ direction.

The in-plane anisotropy $\varepsilon_{ij}\neq 0$  distorts uniform rotation of spins  in space \cite{Nagamiya67,Zhitomirsky96}
\begin{equation}
\theta_i =  {\bf Q}\cdot{\bf r}_i + \varphi_i \ , \ \ \ \varphi_i = \varphi \sin(2{\bf Q}{\bf r}_i)\ .
\label{thetaD}
\end{equation}
Minimization of (\ref{Ecl0}) with respect to $\varphi$ yields to the leading order in small $\varepsilon_{ij}$:
\begin{equation}
\varphi = \frac{2\varepsilon_{\bf Q}} {J_{3\bf Q}-J_{\bf Q}}\ ,
\label{phiD}
\end{equation}
where $\varepsilon_{\bf Q} = \sum_j\varepsilon_{ij}\, e^{i{\bf Qr}_{ij}}$.  Spins bunch towards the easy
direction in the $xy$ plane producing satellite Bragg peaks at ${\bf q} = \pm 3{\bf Q}$ alongside with the
principal peaks at ${\bf q} = \pm {\bf Q}$. The spin bunching also results in an elliptical distortion of the spiral:
\begin{equation}
\frac{\langle S^{x_0}_{\bf Q}\rangle}{\langle S^{y_0}_{\bf Q}\rangle} = 
\frac{1 - d}{1 + d} \ , \ \ \ d = \frac{\varepsilon_{\bf Q}}{J_{3\bf Q}-J_{\bf Q}} \ .
\label{ellip}
\end{equation}
The experimental value  $d \approx 0.13$ for linarite \cite{Willenberg_2012}  can be straightforwardly related 
to the parameters of the model (\ref{Ha}). Using  $\varepsilon_{\bf Q} = - 2\varepsilon |J_1|\cos Q$
and $J_{3\bf Q} -J_{\bf Q} = 8|J_1|\sin^3\!Q\sin 2Q$ with 
the exchange parameters derived by Cemal {\it et al}.~\cite{Cemal_2018} we obtain 
$d \approx 5.6\varepsilon$ that gives an estimate $\varepsilon \sim 0.02$. 
This value of $\varepsilon$ is about 3 times larger than a more precise estimate derived from 
the ESR measurements in Sec.~\ref{sec:ESR}. For more accurate determination of $\varepsilon$ from ellipticity of 
the magnetic structure one may resort to numerical real-space simulations described in \cite{Cemal_2018,Gvozdikova16}.

The excitation spectra are computed in the harmonic approximation neglecting quantum corrections.
For that we use the truncated Holstein-Primakoff transformation for spin components in the local frame:
$S_i^z = S - a_i^\dagger a_i$, $S_i^x \approx\sqrt{S/2}\, (a_i^\dagger + a_i)$, and 
$S_i^y \approx i\sqrt{S/2}\, (a_i^\dagger-a_i)$. Substituting it into Eq.~(\ref{Hloc0}) and keeping only quadratic terms
in boson operators we obtain the harmonic spin-wave Hamiltonian $\hat{\cal H}_2$. After the Fourier transformation
and expansion  in small $\varphi_i$, $\hat{\cal H}_2$ takes the following form:
\begin{eqnarray} 
\hat{\cal H}_2 & = & \sum_{\bf k}  \Bigl[ A_{\bf k} a^\dagger_{\bf k}a_{\bf k}  - 
\frac{1}{2}\, B_{\bf k} (a_{\bf k}a_{-\bf k} + a_{-\bf k}^\dagger a^\dagger_{\bf k} )
\label{Hsw0} \\
& + &   C_{\bf k} (a^\dagger_{\bf k+Q}a_{\bf k-Q} + a^\dagger_{\bf k-Q}a_{\bf k+Q} )  \nonumber \\
&+ & \frac{1}{2}\, D_{\bf k}
(a_{\bf k+Q}a_{\bf -k+Q} + a_{\bf k-Q}a_{\bf -k-Q}  + \textrm{h.c.}) \Bigr] ,
\nonumber 
\end{eqnarray} 
where 
\begin{eqnarray} 
A_{\bf k} & = & \frac{S}{2} J^{zz}_{\bf k} +  \frac{S}{4} (J_{\bf k+Q}+J_{\bf k-Q}) - SJ_{\bf Q} \,, 
\label{ABCD} \\
B_{\bf k} & = & \frac{S}{2} J^{zz}_{\bf k} -  \frac{S}{4} (J_{\bf k+Q}+J_{\bf k-Q}) \,, \ \ 
C_{\bf k}  = D_{\bf k} - \varepsilon_{\bf Q}S \,, \nonumber \\
D_{\bf k} & = & - \frac{S}{4} \Bigl\{\varepsilon_{\bf k} + \varphi\Bigl[ J_{\bf k} -  
\frac{1}{2} (J_{{\bf k}+2{\bf Q}}+J_{{\bf k}-2{\bf Q}})\Bigr] \Bigr\} .
\nonumber
\end{eqnarray} 

The last two terms in (\ref{Hsw0}) vanish for the uniaxial symmetry,  $\varepsilon_{ij}\equiv 0$.
In this case,  $\hat{\cal H}_2$ is diagonalized by the standard Bogolyubov transformation, which eliminates the anomalous terms. The magnon energy is, then,  expressed as
\begin{equation}
\epsilon_{\bf k} = \sqrt{A_{\bf k}^2 - B^2_{\bf k}} 
\label{Ek}
\end{equation}
with  $A_{\bf k}$, $B_{\bf k}$ taken from (\ref{ABCD}).

Using the ESR technique, one can measure magnetic excitations for only a few selected momenta.  
The oscillating radio-frequency (rf) field in ESR experiments is uniform within a  sample: 
\begin{equation}
\hat{V}(t)  = - \sum_i {\bf h}(t)\cdot {\bf S}_i\,.
\label{Vt1}
\end{equation}
Rewritten in the rotating frame (\ref{LF}),  $\hat{V}(t)$ becomes
\begin{equation}
\hat{V}  = -\!\sum_i \bigl\{ h_y(t) S^y_i  +
\bigl[h_x(t)\cos {\bf Qr}_i\! - h_z(t)\sin {\bf Qr}_i\bigr]S^x_i \bigr\},
\label{Vt2}
\end{equation}
where we keep only transverse spin  components and set $\varphi_i=0$ (\ref{thetaD}) for simplicity. 
Equation (\ref{Vt2}) shows that an rf field couples to magnetic excitations with ${\bf k} = 0$ and $\pm{\bf Q}$ and its
polarization determines relative intensity of absorption lines. For a uniform spin spiral,
magnon with ${\bf k} = 0$ has vanishing energy and the resonance spectrum in zero field consists of
two degenerate frequencies corresponding to  ${\bf k} = \pm{\bf Q}$ magnons
\begin{equation}
\Delta_0 = S \sqrt{(J_{\bf Q}^{zz}-J_{\bf Q})\bigl[{\textstyle\frac{1}{2}}(J_{0}+J_{2\bf Q})-J_{\bf Q}\bigr]}\ .
\label{D0}
\end{equation}

Considering now a general case, we note that the additional terms determined by the in-plane anisotropy
mix  a spin wave propagating with momentum $\bf k$ with two other magnons at momenta ${\bf k}\pm 2{\bf Q}$, which in turn are coupled to  excitations with ${\bf k}\pm 4{\bf Q}$ and so on. Calculation of normal modes  requires in the case
diagonalization of infinite matrices  for incommensurate $\bf Q$. 
Following \cite{Zhitomirsky96}, we adopt an approximate method for determining the ESR frequencies.
First, for an incommensurate spiral the ${\bf k} =0$ magnon has zero energy even in the presence of mixing terms. 
The gapless nature of this mode follows from an arbitrary choice of the phase of the incommensurate spiral 
\cite{Elliot66}.  Second, we aim to compute  the ESR gaps or, more precisely, $\Delta^2$ 
including $O(\varepsilon)$ contributions. In this case, the coupling between ${\bf k} = \pm{\bf Q}$  magnons and excitations with ${\bf k} = \pm{3\bf Q}$ can be neglected.
The remaining bosonic terms in (\ref{Hsw0}) are
\begin{eqnarray} 
&&\hat{\cal H}'_2 =  A_{\bf Q} (a^\dagger_{\bf Q} a_{\bf Q}  + a^\dagger_{-\bf Q} a_{-\bf Q}) 
- B_{\bf Q} (a_{\bf Q} a_{\bf -Q} + a^\dagger_{\bf -Q} a^\dagger_{\bf Q} ) 
\nonumber \\
&& \mbox{}+C_0(a^\dagger_{\bf Q} a_{\bf -Q}  + a^\dagger_{-\bf Q} a_{\bf Q}) 
+ \frac{1}{2}D_0
(a_{\bf Q}^2 + a_{\bf -Q}^2  + \textrm{h.c.}).
 \label{Hesr}
\end{eqnarray} 
This quadratic form is diagonalized by introducing symmetric/antisymmetric combinations
$a_{1,2} = (a_{\bf Q}\pm a_{\bf -Q})/\sqrt{2}$, which decouple from each other, and the subsequent Bogolyubov transformation. The obtained energies are
\begin{equation}
\Delta_{1,2}^2 = (A_{\bf Q}\pm C_0)^2 - (B_{\bf Q}\mp D_0)^2 \ .
\end{equation}
Substituting now expressions (\ref{ABCD}) and  keeping only $O(\varepsilon)$ terms we can express the ESR
frequencies as
\begin{eqnarray}
\Delta_{1} & = & S \sqrt{(J_{\bf Q}^{zz}-J^{xx}_{\bf Q})\bigl[{\textstyle\frac{1}{2}}(J_{0}+J_{2\bf Q})-J_{\bf Q}\bigr]}\ ,
\nonumber \\
\Delta_{2} & = & S \sqrt{(J_{\bf Q}^{zz}-J^{yy}_{\bf Q})\bigl[{\textstyle\frac{1}{2}}(J_{0}+J_{2\bf Q})-J_{\bf Q}\bigr]}\ .
\label{D12}
\end{eqnarray}
The splitting between two ESR modes provides a direct measure of the in-plane anisotropy. In the case of linarite
we have
\begin{equation}
\frac{\Delta_1}{\Delta_2} = \sqrt{  \frac{ J_1^{xx} - J_1^{zz}} { J_1^{yy} - J_1^{zz}} }
= \sqrt{  \frac{ \delta + \varepsilon} {  \delta - \varepsilon} } \ ,
\label{ratio}
\end{equation}
which is directly used in Sec.~\ref{sec:ESR} for determining $\delta/\varepsilon$ from the experimental data.

\subsubsection{Finite magnetic fields}

We write the Zeeman energy  in the form
\begin{equation}
\hat{\cal H}_Z = - {\bf H}\cdot \sum_i {\bf S}_i
\label{HZ}
\end{equation}
rescaling magnetic field components with the principal values of the anisotropic $g$ tensor.
By doing that we neglect a possible staggered component of the $g$-tensors of copper ions, which is compatible
with the pleated structure of CuO$_4$ ribbons.
Let us begin with the case of  magnetic field oriented perpendicular   to the helical plane. Spins form  a conical  
structure, which is especially  simple for vanishing in-plane anisotropy: spins tilt towards the field direction 
preserving uniform rotation inside the plane.  The details of corresponding calculations can be found, for example,
in \cite{Zhitomirsky96}. Here we present only the final results.  The magnetic susceptibility per spin is given by
\begin{equation}
\chi_\perp = \frac{1} {(J_0 -J_{\bf Q})} \ .
\label{chi_per}
\end{equation}
Diagonalizing the harmonic spin-wave Hamiltonian one can obtain  an analytic expression
for the  magnon energy in the entire Brillouin zone. One of the ESR frequencies corresponding 
to the ${\bf k} =0$ magnon remains equal to zero, whereas two other gaps are 
\begin{equation}
\Delta_{1,2}= \sqrt{ \Delta_0^2 + H^2\frac{[\frac{1}{2}(J_0\!+\!J_{2\bf Q}) -J_{\bf Q}]^2}{(J_0-J_{\bf Q})^2}}
\pm\frac{H}{2}\frac{J_0-J_{2\bf Q}}{J_0 -J_{\bf Q}},
\label{Dconic}
\end{equation}
where $\Delta_0$ is given by (\ref{D0}).

Magnetic field applied parallel to the helix plane distorts uniform rotation of spins in space:
\begin{equation}
\theta_i =  {\bf Q}\cdot{\bf r}_i -  \alpha_i\ .  
\label{thetaH}
\end{equation}
To the first order in small $H$, the spiral distortion is expressed as \cite{Cooper63, Zhitomirsky96}
\begin{equation}
\alpha_i =  \alpha \sin({\bf Q}{\bf r}_i)   \ , \quad  \alpha = \frac{H}{S[\frac{1}{2}(J_0+J_{2\bf Q}) -J_{\bf Q}]} \ . 
\label{alpha}
\end{equation}
Accordingly,  the magnetic susceptibility is 
\begin{equation}
\chi_\parallel = \frac{1} {(J_0 + J_{2\bf Q} - 2J_{\bf Q})} \ .
\label{chi_par}
\end{equation}
An important characteristic of a spin helix is anisotropy of the susceptibility tensor 
$\chi_\perp/\chi_\parallel$ or  
\begin{equation}
\eta = \frac{\chi_\perp-\chi_\parallel}{\chi_\parallel} = 
\frac{J_{2\bf Q} - J_{\bf Q}}{J_0  - J_{\bf Q}} \ .
\label{eta}
\end{equation}
Generally, one has $\eta>0$ or $\chi_\perp/\chi_\parallel>1$, since $\bf Q$ corresponds 
to the minimum of the Fourier transform $J_{\bf q}$ (\ref{Jq}).
In particular, this means that in the absence of anisotropy, the helical plane is oriented perpendicular
to the applied magnetic field. Values of $\chi_\perp/\chi_\parallel$ computed for different sets 
of the exchange parameters are shown in Table~\ref{parameters} \cite{remark}.

Calculation of the ESR frequencies from the spin-wave theory becomes rather cumbersome  for
magnetic field applied in the plane of spin helix \cite{Cooper63}.  
Instead, in the next subsection we present a simple derivation of the corresponding 
ESR spectra within the framework of the macroscopic field-theoretical approach.

The special feature of the phase diagram of linarite is presence of 
a commensurate canted antiferromagnetic state for a magnetic field applied parallel to the  
intermediate $y$ axis \cite{Willenberg_2016}. This two-sublattice state is described by the propagation vector
${\bf Q}_0 = (0,0,\pi)$ and appears due competition between incommensuration 
and the in-plane anisotropy \cite{Cemal_2018}. Computation of the excitation spectra within 
the harmonic spin-wave theory is straightforward  for this state. The two ESR frequencies corresponding to
magnons with ${\bf k}=0$ and ${\bf k}={\bf Q}_0$ are
\begin{eqnarray}
\Delta_{1} & = & 2|J_1|S \sqrt{(2j_c+\delta+\varepsilon) (2j_c\sin^2\beta +\delta\cos^2\beta)} \ ,
\nonumber \\
\Delta_{2} & = & 2|J_1|S \sqrt{(\delta+\varepsilon) (2j_c +\delta)}\cos\beta \ ,
\label{Dcom}
\end{eqnarray}
where $j_c = J_c/|J_1|$ and $\beta$ is the canting angle:
$$
\sin\beta = \frac{H}{2S|J_1|(\delta+2j_c)}\,.
$$
Comparison of the experimental values of the ESR gap in the commensurate state
 to the theoretical expressions  (\ref{Dcom}) provides another consistency check  for different sets of microscopic
 parameters for linarite.

\subsection{Macroscopic Theory}
\label{sec:macroTh}

An alternative prospective on the ESR spectra of ordered and disordered magnets is provided by the macroscopic 
or hydrodynamic theory of magnetization dynamics. This field-theoretical approach was developed  by 
several authors \cite{Halperin69,Halperin77,Andreev80,Affleck89}, who applied it to both ordered 
and disordered magnetic phases.
The main idea is to  `integrate out' the high-energy excitations and focus exclusively on the long-wavelength, 
low-energy modes. The long-wavelength oscillations only weakly perturb an underlying magnetic structure. Therefore, one can introduce a local antiferromagnetic  order parameter and investigate its dynamics using the gradient expansion. For dominant exchange interactions, the emergent energy functional or Lagrangian has a universal form. For a collinear Heisenberg antiferromagnet it coincides with that for the nonlinear $\sigma$-model \cite{Andreev80,Affleck89}. 
In the subsequent analysis we follow the formulation by Andreev and Marchenko \cite{Andreev80}, who specifically
considered the ESR dynamics of ordered antiferromagnets.

The staggered magnetization of a spiral antiferromagnet formed by competing exchange interactions 
is described by a pair of orthogonal unit vectors ${\bf l}_1$ and  ${\bf l}_2$ that determine the helix plane:
\begin{equation}
\langle {\bf S}_i\rangle \simeq {\bf l}_1 \cos{\bf Qr}_i + {\bf l}_2 \sin{\bf Qr}_i \ .
\end{equation}
In addition, we define the vector ${\bf l}_3 = {\bf l}_1\times {\bf l}_2$ normal to the spin plane. 
In the uniform state, the Lagrangian density satisfying all symmetry requirements is
\begin{equation}
{\cal L} = \sum_{k=1}^3 \frac{\chi_k}{2\gamma^2}\, \bigl[\partial_t {\bf l}_k - \gamma ({\bf l}_k\times{\bf H})\bigr]^2 - E_a \ ,
\label{L}
\end{equation}
where $\gamma = g\mu_B/\hbar$ is the gyromagnetic ratio and $E_a$ is the anisotropy energy. Note, that the combination of
time derivatives and the applied magnetic field in (\ref{L}) is uniquely determined by the Larmor theorem \cite{Slichter}.
Considering the static case, one can relate the constants $\chi_k$ with the components of the susceptibility tensor:
\begin{equation}
\chi_1 = \chi_2 = {\textstyle \frac{1}{2}}\chi_\perp \ , \quad \chi_3 = \chi_\parallel - {\textstyle \frac{1}{2}}\chi_\perp 
\end{equation}
with $\chi_\perp$ and $\chi_\parallel$ given by Eqs.~(\ref{chi_per}) and (\ref{chi_par}), respectively.
At the next step,  oscillations of the triad ${\bf l}_k(t)$ around the equilibrium position have to be parameterized 
by three angles and the full expression (\ref{L}) can be expanded to the second order in small deviations. We skip 
technical details  referring instead to  the previous
applications of the macroscopic theory  for planar spiral antiferromagnets 
see \cite{Zaliznyak88, Abarzhi93, Svistov09}. 

In the case of linarite, the biaxial anisotropy energy can be represented in terms
of the components of the perpendicular vector ${\bf l}_3$:
\begin{equation}
E_a  = -\frac{D}{2}\, l_{3z}^2 - \frac{E}{2} \bigl(l_{3y}^2  - l_{3x}^2 \bigr)\,,
\label{Ea}
\end{equation}
where $D>0$ determines the orientation of the spiral plane in zero field 
and $E>0$ selects the easy axis direction inside the plane. Note that the signs of anisotropy constants 
in (\ref{Ea}) are opposite to that of actual spins.
For an incommensurate spiral the anisotropy energy $E_a $ does not contain locking terms
that describe preferable orientation of the pair $({\bf l}_1,{\bf l}_2)$  inside the helix plane.  However,
such pinning potential emerges for  commensurate structures, see an example of the six-sublattice triangular antiferromagnet discussed in \cite{Abarzhi93}.
Thus, we conclude that for linarite one  of the ESR  modes always has zero frequency irrespective
of the applied magnetic field. 

In zero field, the two nonzero ESR frequencies are 
\begin{equation}
\frac{\omega_{1,2}^2}{\gamma^2} =\frac{D\pm E}{\chi_\parallel} \ .
\label{gap}
\end{equation}
Comparing these with Eq.~(\ref{D12}) we relate the phenomenological constants with 
the microscopic parameters:
\begin{equation}
D = S^2\bigl[J_{\bf Q}^{zz} - {\textstyle \frac{1}{2}}(J_{\bf Q}^{xx} + J_{\bf Q}^{yy})\bigr], \quad
E = \frac{S^2}{2}(J_{\bf Q}^{yy} - J_{\bf Q}^{xx}).
\label{DE}
\end{equation}

For the uniaxial anisotropy ($E=0$) and magnetic field oriented perpendicular to the helix plane 
one finds
\begin{equation}
\frac{\omega_{1,2}}{\gamma} = \sqrt{ \frac{D}{\chi_\parallel} + \frac{(1+\eta)^2}{4}H^2 } 
\pm \frac{1-\eta}{2}\, H  \ ,
\end{equation}
where, as before, $\eta = (\chi_\perp-\chi_\parallel)/\chi_\parallel$.
The above expression  fully agrees with the spin-wave result (\ref{Dconic}) confirming the equivalence
between microscopic and macroscopic approaches. 

For magnetic field applied parallel to the helix plane, the macroscopic theory gives a simple result
\begin{equation}
\frac{\omega^2_1} {\gamma^2} = \frac{D}{\chi_\parallel}\,, \quad
\frac{\omega^2_2} {\gamma^2}=\frac{D}{\chi_\parallel} + H^2\,. 
\label{Dtransv}
\end{equation}
Note, that the corresponding spin-wave expressions are obtained only after a lengthy and much more involved calculation
\cite{Cooper63}. The above expression is valid for magnetic fields that do not exceed the spin-flop field
\begin{equation}
H_{\rm sf} = \sqrt{\frac{D}{\chi_\perp - \chi_\parallel}} \ .
\label{HsfD}
\end{equation}
Above  $H_{\rm sf}$ the spiral plane changes its orientation to orthogonal with respect to the field.

For the biaxial magnetic anisotropy the resonance frequencies in the orthogonal geometry ${\bf H}\parallel\hat{\bf z}$ 
are given by 
\begin{eqnarray}
\label{Dperp}
\frac{\omega_{1,2}^2}{\gamma^2} & = & \frac{1+\eta^2}{2} H^2 + \frac{D}{\chi_\parallel} 
\\
& \pm & \sqrt{  \frac{E^2}{\chi_\parallel^2} +  \frac{D}{\chi_\parallel} H^2(1-\eta)^2 + \frac{(1-\eta^2)^2}{4} H^4}\,.
\nonumber
\end{eqnarray}
For fields parallel to the spiral plane, the spin-flop transition  becomes orientation dependent. 
For the two principal directions one finds
\begin{equation}
H_{\rm sf}^{x,y} = \sqrt{\frac{D\pm E}{\chi_\perp - \chi_\parallel}} \ .
\label{HsfDE}
\end{equation}
The ESR frequencies  are described by expressions that are similar to Eq.~(\ref{Dtransv}):
\begin{equation}
\frac{\omega^2_1} {\gamma^2} = \frac{D\mp E}{\chi_\parallel}\,, \quad
\frac{\omega^2_2} {\gamma^2}=\frac{D\pm E}{\chi_\parallel} + H^2\,, 
\label{DtransvDE}
\end{equation}
where the upper  and the lower signs correspond to ${\bf H}\parallel\hat{\bf x}$ and ${\bf H}\parallel\hat{\bf y}$,
respectively.

\section{Experiment}
\label{sec:exp}

\subsection{Technical Details} 

In our experiments we have measured a naturally grown single crystal of linarite from 
the Grand Reef Mine, Arizona, USA. The crystal is from the same batch as the one used in 
the previous studies of dielectric and thermodynamic properties \cite{Povarov_2016, Feng18}. 
The crystal has a shape  of a prism with dimensions $2.7\times 0.8\times 0.8$~mm$^3$. 
The extended edge is parallel the $b$ axis, allowing straightforward orientation with respect to 
this crystallographic direction. 

Our ESR setup is equipped with a multiple mode resonator of the transmission type in the frequency range 
$18 < \nu < 140$~GHz and magnetic field up to 9~T. The sample was glued on a rotating holder.
During experiment, temperature was varied between $0.5$ and $25$~K. Measurements in the high-frequency range 
$150-250$~GHz have been conducted using the quasioptical terahertz spectroscopy, see e.g.
\cite{Kuzmenko16}, in magnetic fields up to 7~T at $T = 1.8$~K.
All temperature values were stabilized with precision better than $0.05$~K. 

The orientation of the $x$ and $z$ axes have been identified in the ESR angular dependence measurements using the fact that the static magnetic field applied along the $x$ axis causes the spin-flop reorientation at $\mu_0 H_{\rm sf}^x=3$~T \cite{Cemal_2018}. This anomaly is easily detectable by a step-like anomaly in the field scans of transmitted power in ESR experiments. This
feature allows to identify the $x$ axis with precision better than $5^\circ$.

\subsection{ESR Results}
\label{sec:ESR}

\begin{figure}
\includegraphics[width=0.85\columnwidth]{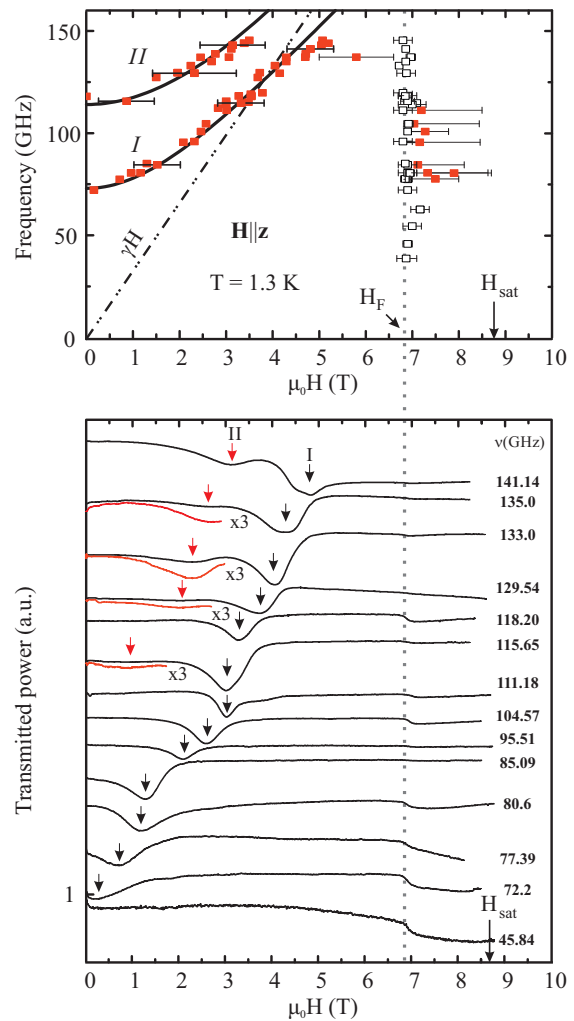}
\caption{(Color online.) The frequency--field diagram for $\vect{H}\parallel \vect{z}$ measured at $T=1.3$~K. The ESR resonances are marked by solid symbols, empty squares correspond to phase transitions observed as anomalies on $P_{tr}(H)$ scans. The dotted line shows the transition field into the fan state. Solid lines are theoretical curves $\nu(H)$ computed for the spiral state. Dash-dotted line is the paramagnetic $\nu(H)$. 
Bottom panel: examples of $P_{tr}(H)$-scans measured at $T=1.3$~K.
}
\label{Hparc}
\end{figure}

Our main results are summarized in the frequency--field ($\nu$--$H$) diagrams, Figures~\ref{Hparc}, \ref{Hparb},
and \ref{Hpara}, which show the frequency dependence of resonance fields for the three principal 
field orientations. 
Two excitation branches with distinct zero field gaps are observed in each case providing clear evidence
for a substantial biaxial anisotropy in the system. 
The field behavior of the ESR gaps allows to distinguish different magnetic structure identified for linarite in the neutron diffraction experiments, see Fig.~\ref{structurenew}. Detailed comparison with theoretical predictions is performed for the
cycloidal and the conical spirals as well as for the field-induced commensurate antiferromagnetic state. In the narrow field region  occupied by the fan phase, the resonance lines substantially broaden and do not allow for a meaningful comparison with the theory.

\subsubsection{Magnetic field parallel to the $z$-axis}

Figure \ref{Hparc} shows the frequency-field diagram together with examples of the field scans of transmitted through the resonator high-frequency power ($P_{tr}(H)$) measured at $T=1.3$~K for applied field oriented perpendicular to the spiral plane: $\vect{H} \parallel \vect{z}$. In this orientation,
a conical or an umbrella  state is realized in a wide range of fields up to the transition ($H_F$) into the fan phase  in the close vicinity of the  uniformly magnetized ``saturated" phase (see bottom panel of  Fig.~\ref{structurenew}). Transition to the fan phase is detected on $P_{tr}(H)$-scans as a step-like deflection. The singularity fields measured at different frequencies are shown at the frequency--field diagram with open symbols. The obtained value of $H_F$ is in full agreement with the values reported in Ref.~\cite{Cemal_2018}.

\begin{figure}[t]
\includegraphics[width=0.85\columnwidth]{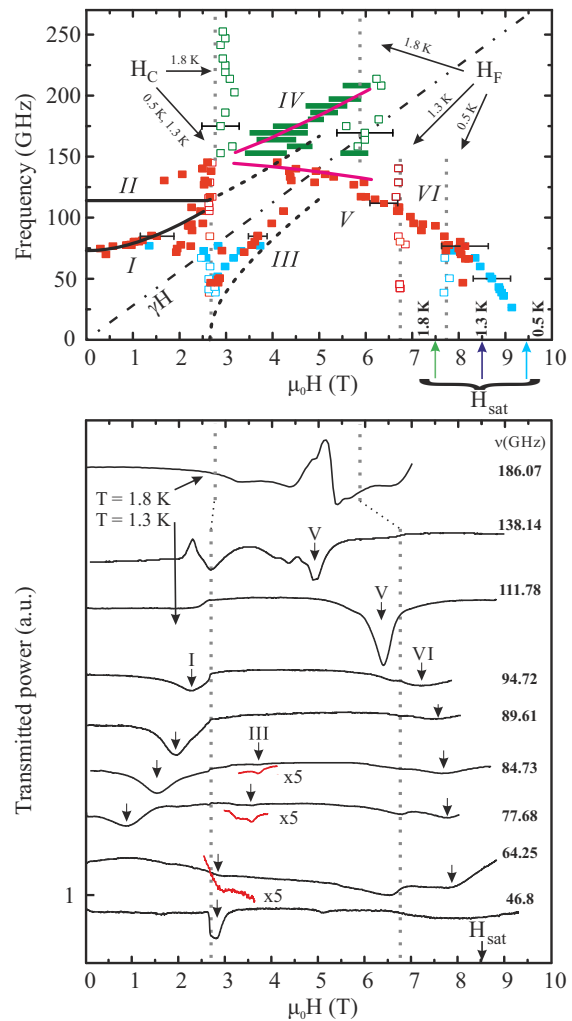}
\caption{(Color online.) Upper panel: the frequency--field diagram for $\vect{H}\parallel \vect{y}$. Resonances corresponding to absorption lines measured at different temperatures are shown by symbols of different colors (blue squares -- $T=0.5$~K, red squares -- $T = 1.3$~K, green rectangles -- $T=1.8$~K). Empty squares are measured transition fields. The vertical dotted lines show transition fields from the bulk measurements \cite{Povarov_2016}
obtained for $T=0.5$~K ($\mu_0 H_C = 2,7$~T, $\mu_0 H_F = 8$~T), $T = 1.3$~K ($\mu_0 H_C = 2,7$~T, $\mu_0 H_F = 6.7$~T) and $T=1.8$~K ($\mu_0 H_C = 2.8$~T, $\mu_0 H_F = 6$~T). Solid black and purple lines show theoretical curves $\nu(H)$ computed for spiral and  canted antiferromagnetic phases. Dash-dotted line is the paramagnetic line.
Bottom panel: examples of $P_{tr}(H)$. Upper line was measured at $T=1.8$~K, other lines -- at $T=1.3$~K.}
\label{Hparb}
\end{figure}

The resonance fields measured at different frequencies are shown  by solid symbols. The diagram features two ESR gaps  in the spiral phase  denoted as `I' and `II', which rise with increasing
magnetic field. The error bars for some of the points illustrate characteristic values for the width at the half-height of 
the absorption lines. 
The asymptotic slope of low-frequency branch `I', $\nu_1 \to \gamma\eta H$, determined by the susceptibility anisotropy of  the spin spiral  is smaller than the inclination of the paramagnetic resonance $\nu/H = \gamma$  shown by the  dash-dotted line. Fitting the experimental data with the theoretical formula Eq.~(\ref{Dperp}) is shown by solid lines. The parameters obtained from these fits are $\nu_{10}=73 \pm 1$~GHz, $\nu_{20} =114 \pm 2$~GHz, 
and $\chi_\perp/\chi_\parallel=1.85 \pm 0.1$. The $g$-factor values used in our computations, $g_y = 2.1$, 
$g_{x}\approx g_z = 2.3$, are taken from \cite{Wolter_2012}.
The same parameters are also used for theoretical fits in other field geometries, Figs.~\ref{Hparb}--\ref{anglebc}.
The value of $\chi_{\perp}/\chi_{\parallel} = 1.85 \pm 0.1$  agrees with the result from static magnetization measurements $\chi_{\perp}/\chi_{\parallel}=1.6\pm 0.1$~\cite{Yasui_2011}, which were 
performed at somewhat higher $T=2$~K and, thus, are further  away from the zero temperature
limit. A brief comparison of the obtained result and theoretical values  derived for several sets
 of microscopic exchange parameters  is deferred to Sec.~IV.

\subsubsection{Magnetic field parallel to the $y$-axis}

The upper panel of Fig.~\ref{Hparb} presents the $\nu$--$H$ diagram for  $\vect{H}\parallel \vect{y}$ combining
data from several temperatures: $T=0.5$~K, $T=1.3$~K, and $T=1.8$~K . 
For illustration we show on the lower panel examples of the power absorption scans $P_{tr}(H)$. 
Anomalies corresponding to the phase transitions from the spiral to the commensurate
antiferromagnetic phase and from the antiferromagnetic phase to the fan phase are well detected 
on $P_{tr}(H)$-scans. The anomaly fields measured at different frequencies are shown by empty symbols. 
The transition fields exhibit significant temperature dependence. 
Summarizing the data observed at different frequencies and temperatures, 
the field ranges of transitions are marked by vertical lines. 
The transition from the fan to saturated phase shows no deflections on 
the $P_{tr}(H)$-scans. Values of the saturation field shown in the Figure are 
taken from Ref.~\cite{Povarov_2016}.

The resonance fields obtained at different frequencies are shown on the diagram with solid squares. 
The error bars for selected points indicate the characteristic width at the half-height of the absorption lines. 
The resonance lines measured with the quasioptical technique have complicated shapes because the transmission of 
high-frequency (HF) power through a small diaphragm with the linarite sample of a nonregular shape depends not only 
on the real part of the HF-susceptibility but also on the imaginary part. Elongated shaded rectangles on the $\nu$--$H$ diagram corresponding to this frequency range mark the full area of peculiarities on $P_{tr}(H)$ scans. 

In the  low-field region $H<H_C$ linarite has the incommensurate spiral magnetic structure \cite{Willenberg_2016}.  For $\vect{H} \parallel \vect{y}$ only one field-increasing branch  was detected marked as `I' on the diagram. 
This branch has the energy gap $\nu_{10} = 73$~GHz. Branch `II' corresponding to $\nu_{20} = 114$~GHz is field independent for this orientation, which precludes its observation in the field-scan measurements. Nevertheless, this branch can be detected for the two other field directions $\vect{H}\parallel\vect{x},\vect{z}$.

In the field region $H_C < H <  H_F$ with the canted antiferromagnetic structure the two  intensive resonance 
absorption lines have been detected: the field increasing branch `IV' and the declining branch `V'. We ascribe these branches to oscillations  with the wave vectors $\vect{k} = (0,0,0)$ and $(0,0,\pi)$, specific to the commensurate 
two-sublattice  structure. The solid red lines show the theoretical spectra Eq.~(\ref{Dcom}) computed using  
 $J_1 = -14.5$~meV and $J_c= 0.7$~meV, $\delta/\varepsilon = 2.4$ obtained previously from 
 the ESR gaps in zero field and adjusting the remaining parameter as $\varepsilon= 0.006$.
For these fits we have chosen the set of exchange parameters obtained from the INS measurements 
in the saturated phase \cite{Cemal_2018}, since these parameters are not affected by quantum fluctuations.

\begin{figure}[t]
\includegraphics[width=0.85\columnwidth]{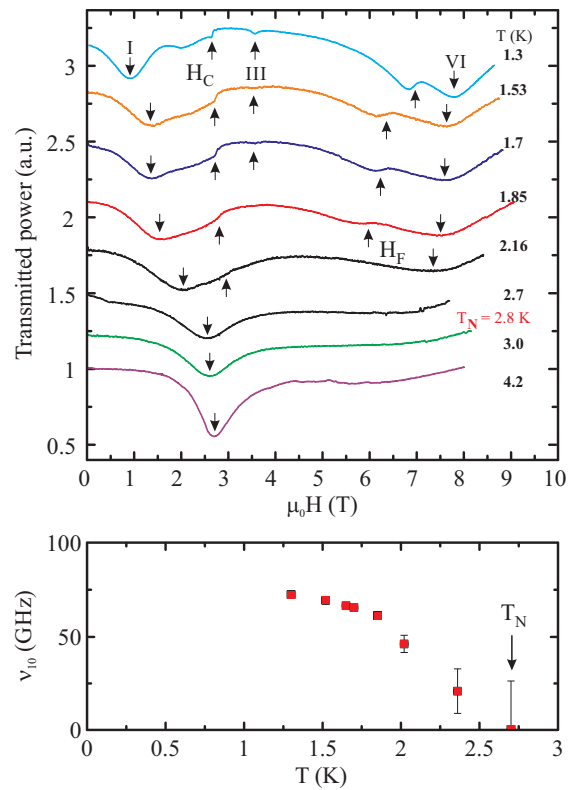}
\caption{(Color online.) Upper panel: The temperature evolution of the power absorption scans $P_{tr}(H)$ 
measured at $\nu = 77.72$~GHz for $\vect{H} \parallel \vect{y}$. The roman numbers correspond to the notations of branches  in Figs.~\ref{structurenew} and \ref{Hparb}.
Bottom panel: temperature dependence of the lowest zero-field gap $\Delta_{1}$.
}
\label{HTb}
\end{figure}

In addition to these resonances, a narrow low-intensity absorption line (branch `III') was 
also detected for $H> H_C$. Its intensity decreases with the increasing field (frequency) and 
vanishes at $\mu_0 H\sim 4$~T. The additional branch can be ascribed to oscillations of 
{\it spiral} spin-floped structure. We suggest that the branch `III' corresponds to the absorption 
in the part of our sample which continues to be in the spiral phase after the spin plane 
reorientation. Indeed, the computed value of the spin-flop transition field for 
$\vect{H} \parallel \vect{y}$ is equal to $\mu_0 H_{\rm sf}^y = 2.8~\pm~0.2$~T, see 
Eqs.~(\ref{gap}) and (\ref{HsfDE}), which is somewhat larger but close to the transition into 
the canted commensurate phase $\mu_0H_C=2.7$~T realized  experimentally. Theoretical dependence 
$\nu(H)$ for the spiral phase above the spin-flop transition $H>H_{\rm sf}^y$ is shown in 
Fig.~\ref{Hparb} with the dashed line. The angular dependence of the resonance field of the 
line `III' is also satisfactorily explained by oscillations of spiral spin structure (see 
Fig.~\ref{anglebc}). For this reason we suppose that the transition to the canted antiferromangetic phase is 
broad and in the field range $2.5$~T$<\mu_0 H<4.5$~T commensurate and spin-floped spiral phases 
coexist. Coexistence of the low-field incommensurate spiral and the high-field 
commensurate antiferromagnetic phase  was also found in several previous studies
and was assigned to the region II on the phase diagram of Ref.~\cite{Willenberg_2016}.

The field-decreasing branch `V' detected in the canted commensurate phase continues smoothly into 
a much broader line `VI' in the high-field region. The similarity of the ESR spectra 
in the two phases yields further support to the presence of the fan phase near $H_{\rm sat}$ \cite{Cemal_2018}.
The other suggestion for the high-field region advocated in literature is 
the longitudinal spin-density wave  (SDW)  state \cite{Willenberg_2016,Heinze19}. Indeed,
the SDW state was found in another $J_1$--$J_2$ chain material LiCuVO$_4$ at {\it intermediate} fields. 
However, it exhibits quite different ESR spectra from what we observe in linarite \cite{Prozorova16}.
Resonance frequencies measured at $T=0.5$~K (blue squares at the $\nu$--$H$ diagram) are extrapolated 
to zero  at the saturation field $H_{\rm sat}$.
 
Figure~\ref{HTb} shows the temperature evolution of $P_{tr}(H)$-scans measured at $\vect{H} \parallel \vect{y}$, $\nu = 77.72$~GHz. The $P_{tr}(H)$-scans are normalized to $1$ at $H=0$ and shifted along the ordinate axis for visibility. The shift of singularities with temperature is in good agreement with transition lines between spiral, canted commensurate, and  fan phases of the $H$--$T$ diagram obtained from the bulk measurements in \cite{Povarov_2016}. The shift of the resonance line `I' to higher fields with temperature can be explained by temperature decrease of the energy gap 
$\nu_{10}$. The bottom panel of Fig.~\ref{HTb} illustrates the temperature dependence of the energy gap $\nu_{10}$.
The gap vanishes in a close vicinity of the transition temperature $T_N=2.8$~K.

\subsubsection{Magnetic field parallel to the $x$-axis}
\label{sec:ESRx}

\begin{figure}[t]
\includegraphics[width=0.85\columnwidth]{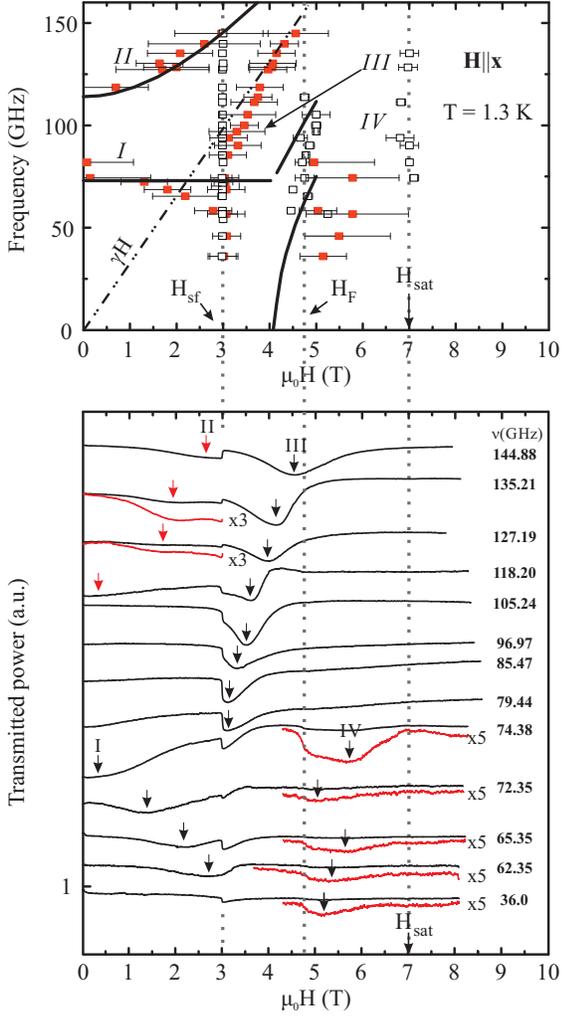}
\caption{(Color online.) 
Upper panel: the frequency--field diagram for $\vect{H}\parallel \vect{x}$. 
The solid lines show theoretical $\nu(H)$ computed for the spiral phase before and after the spin-flop transition.
Dashed lines indicate the transition fields. Dash-dotted line gives the paramagnetic $\nu(H)$.
Bottom panel: examples of $P_{tr}(H)$ recorded at $T=1.3$~K.
}
\label{Hpara}
\end{figure}

The upper panel of Fig.~\ref{Hpara} shows the frequency--field diagram measured at $T=1.3$~K for  $\vect{H}\parallel \vect{x}$. The power transmission field scans $P_{tr}(H)$ allow to observe anomalies corresponding to the spin-flop reorientation as well as to the phase transitions between the spiral and the fan phases and between the fan phase and the saturated state.
The two ESR branches, marked as `I' and `II,' are distinguished in the spiral phase  for fields below the spin-flop transition. The lower branch `I' corresponds to oscillations of the spiral plane around 
the $x$-axis and is expected to be independent of the applied field, see Sec.~II. 
Nevertheless, a weak decrease of the corresponding ESR gap with magnetic field is clearly seen in the experimental data. We exclude a small field misorientation as a possible reason for that since the spin-flop transition remains quite sharp. 

The discrepancy with
the theory may be caused either by strong distortions of the spin structure induced
by the field applied in the spiral plane or by deviations from the minimal spin model.
In particular, our spin model does not include a staggered component of the $g$-tensors of copper ions 
and the Dzyaloshinskii-Moriya interaction on the nearest-neighbor bonds. 
These may be also the reason for the discrepancy between the calculated  $\mu_0 H_{\rm sf}^x=4$~T and the experimental $\mu_0 H_{\rm sf}^x=3$~T values of the spin-flop transition field into the conical state.
In the region of the fan phase a
broad resonance absorption line has been observed marked as `IV' in the $\nu$--$H$ diagram.

\subsubsection{Magnetic field within $xz$- and $yz$-planes}

\begin{figure}[b]
\includegraphics[width=0.95\columnwidth]{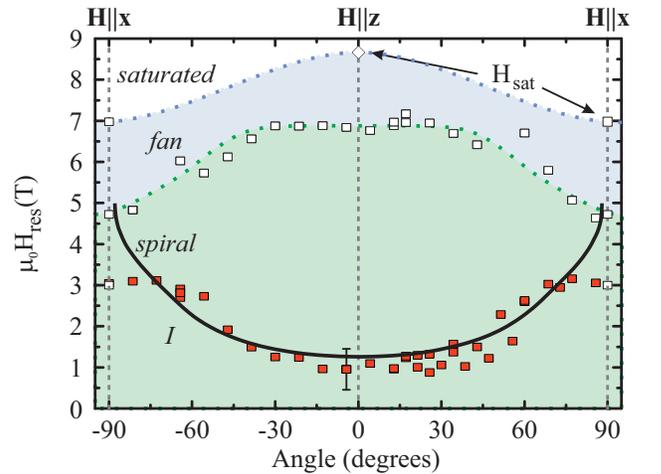}
\caption{(Color online.) Angular dependence in the $xz$ plane of the resonance field  for $\nu=80.6$~GHz, solid squares,  and the phase transition fields, empty symbols,  measured at $T=1.3$~K. 
The solid line is a theoretical curve computed for the spiral magnetic structure. Dotted lines are guides for the eye. 
}
\label{angleac}
\end{figure}

\begin{figure}[t]
\includegraphics[width=0.95\columnwidth]{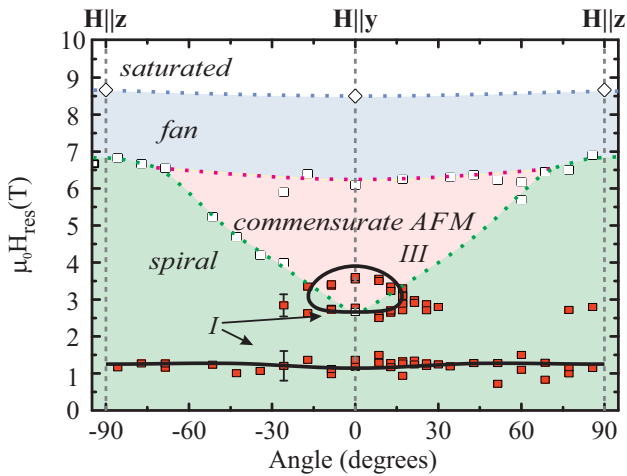}
\caption{
(Color online.) 
Angular dependence  in the $yz$ plane of the resonance field  for $\nu=80.6$~GHz, solid squares,  
and the phase transition fields, empty symbols,  measured at $T=1.3$~K. The solid lines are theoretical curves computed for the spiral magnetic structure. Dotted lines are guides for the eye. 
}
\label{anglebc}
\end{figure}

The angular variations of the resonance field for $\nu=80.6$~GHz and the phase transition fields are presented 
in Figs.~\ref{angleac} and \ref{anglebc} for magnetic fields rotated in the $xz$- and the $yz$-planes, respectively. 
The data have been collected at $T=1.3$~K. 
 The transition from the fan to the saturated phase is added to the figure from the $\nu$--$H$ diagram measured at $\vect{H} \parallel \vect{x}$ (Fig.~\ref{Hpara}). Transition fields $H_{sat}$ for  $\vect{H} \parallel \vect{y}$ and $\vect{z}$ marked in the figures by diamonds, are taken from Refs.~\cite{Povarov_2016, Cemal_2018}. The solid lines in the angular dependences show the computed dependences of resonances corresponding to branches `I' and `III' in spiral phase.
The theoretical curves are in agreement with experimental points in low field range. At higher fields the discrepancy increases, probably due to essential distortion of the spin structure. Note here, that the additional branch `III' is satisfactorily fitted by the spectrum of oscillations of the spin-floped spiral state inside the field region, where the canted antiferromagnetic state  is stable. As we already remarked before, this observation fully  agrees  with the conclusion reached in the previous works about coexistence of commensurate and spin-floped spiral structures at fields above $H_C$ at $\vect{H} \parallel \vect{y}$ \cite{Willenberg_2016, Povarov_2016}.

\section{Summary and discussion}

We begin by summarizing  the magnetic parameters of linarite extracted from theoretical fits
of the ESR spectra. The spiral magnetic structure  is characterized  in zero field by two 
susceptibilities: parallel $\chi_{\parallel}$ and perpendicular $\chi_\perp$ to the spiral plane 
with $\chi_{\perp}/\chi_{\parallel} = 1.85~\pm~0.1$. This value is compared in Table I to the classical 
theory result (\ref{eta}) computed for various sets of the exchange parameters  and 
with the dc magnetization measurements of Yasui {\it et al.}~\cite{Yasui_2011}. Note, that our experiments
were performed at $T=1.3$~K whereas  results  reported in \cite{Yasui_2011}
correspond to higher $T=2$~K, which explains difference  between the two values.   
Comparison with theoretical results clearly shows significance of interchain coupling for linarite and inadequacy of 
modelling its properties with the purely one-dimensional spin Hamiltonians. 
Apart from that, the three sets of microscopic couplings predict  $\chi_{\perp}/\chi_{\parallel}$ fairly close to our
value.  The most precise values of the exchange constants are expected to be determined by the INS measurements 
in the saturated phase \cite{Cemal_2018}, see Table I. We ascribe the remaining 15--20\%\ difference between 
the experimental and theoretical results for $\chi_{\perp}/\chi_{\parallel}$
to the quantum renormalization  effect. Indeed, quantum fluctuations  in linarite are non-negligible 
due to the dominant intrachain interactions and the discussed 
effect for  $\chi_{\perp}/\chi_{\parallel}$ roughly matches the observed reduction of
ordered moments in zero field \cite{Willenberg_2012}.

Based on the experimental values of zero-field ESR gaps  and using 
$\chi_\perp = 0.068 \pm 0.004\ \mu_B$/Cu/T$ = 4900 \pm 300$~J/T$^2$/m$^3$ \cite{Cemal_2018} 
we have obtained for the macroscopic anisotropy parameters of linarite as defined in Eq.~(\ref{Ea}): 
$D = 30~\pm~2$~kJ/m$^3$ and $E = 12.6~\pm~1$~kJ/m$^3$.  The biaxial anisotropy is significant 
and, as a result, the spin-flop transition depends on the field orientation inside the helix plane. 
Furthermore, using the minimal microscopic model of the biaxial anisotropy (\ref{Ha}) with the exchange
constants of Cemal {\it et al.}\  \cite{Cemal_2018} we have 
determined  the exchange anisotropy parameters $\delta \approx 0.014$ and $\varepsilon\approx 0.006$. 
These values agree with the general result that  symmetric anisotropic interactions 
constitute  $(\Delta g/g)^2$ fraction of the isotropic exchange with  $\Delta g$ being the anisotropic part
of the $g$ factor of a magnetic ion  \cite{Moriya60}. The obtained values for $\delta$ and $\varepsilon$ 
are more accurate than the previous estimates  $\delta \sim 0.03$ and 
$\varepsilon\lesssim 0.003$ \cite{Cemal_2018}, since our results have been derived by directly measuring 
the small excitation gaps. The anisotropy parameters found in our study provide
a reference point for future theoretical work on linarite.

Despite good overall agreement, there is one inconsistency between the ESR theory of a spiral
antiferromagnet  and the experimental behavior of resonance frequencies for ${\bf H}\parallel{\bf x}$,
see Sec.~\ref{sec:ESRx} and Fig.~\ref{Hpara}. The disagreement is probably
due to extra anisotropic terms besides those included in Eq.~(\ref{Ea}).
At the microscopic level extra terms can be generated either by the staggered DM interaction on the $J_1$ bonds
or by staggered component of the $g$ tensors of copper ions both excluded in the minimal model (\ref{Ha}).
The latter interaction is probably more important since the disagreement between theory and experiments
builds on with increasing magnetic field. This problem needs to be addressed in future  
studies on linarite. 

Another point deserving special comment is why 
the semiclassical theory of incommensurate magnetic spirals works so well for linarite,
though the material is obviously quasi one-dimensional, see Table I. We attribute this fact to
well developed ordered moments found in the neutron diffraction experiments \cite{Willenberg_2012}. 
The related suppression of quantum fluctuations is a combined effect of magnetic anisotropy and interchain exchange coupling. Still, according to the published zero-field data \cite{Rule_2017}, the high-energy dynamics of linarite appears to be quite unusual. In particular, no excitations have been detected above 1.5 meV despite the fact that the magnon band computed within the harmonic spin-wave theory extends up to $\sim 20$~meV. Hence, the spin dynamics of linarite 
is far from being trivial and deserves additional experimental and theoretical investigation.

Regarding exotic multipolar quantum states argued to be realized in linarite \cite{Willenberg_2016},
our measurements do not allow for direct verification of their presence or absence.
We refer to the recent NMR work \cite{Heinze19}, which leaves only a narrow window of fields 9.35 T$ <\mu_0 H < 9.64$~T for a possible multipolar state in the ${\bf H}\parallel{\bf b}$ geometry. If continuous spin rotations 
about the field direction are present, a multipolar state is formed by condensate of bound complexes formed
by $m$ spin flips. Since the total spin projection $S^z_{\rm tot}$ is a good quantum number in the saturated phase, each $m$ sector appears to be independent and the one with the highest critical field determines
the equilibrium multipolar state. The multipolar states considered for linarite include 
quadrupolar or spin-nematic states with $m=2$, octupolar states with $m=3$ and other states up to $m=6$ 
\cite{Willenberg_2016}. However, the above classification of multipolar states fails if continuous symmetry
is replaced with only discrete $p$-fold rotations about the external field direction. In this case one can characterize the magnon condensates by$\mod(m,p)$ only. The observed biaxial anisotropy in linarite
leaves two possibilities: a multipolar
state can be described  either by $m=1$ and, thus, have the trivial dipolar symmetry or by $m=2$
corresponding to a nontrivial quadrupolar state. The quadrupolar state can additionally break the translation symmetry
producing spontaneous dimerization \cite{Zhang17}.
Whether such a state is realized in linarite in the narrow high-field region with ${\bf H}\parallel{\bf b}$  
remains to be seen.

\section{Acknowledgements}

We are grateful to K.~Yu.~Povarov and A.~Zheludev for providing  the linarite crystal used in our measurements
and for helpful discussions of the results. 
We also thank  M. Enderle and B. F\aa k for useful comments on the manuscript.
The work was supported in part by the Program of the Presidium of RAS 1.4 ``Actual problems of low temperature physics", by Russian Foundation for Basic Research (Grant No.\ 19-02-00194), by the Austrian
Science Funds (Grant No.\ I 2816-N27), and by the ANR  project Matadire. The low-temperature ESR measurements 
at $T = 0.5$~K were supported by Russian Science Foundation Grant No.\ 17-02-01505.


\end{document}